\documentclass[prl,twocolumn,showpacs,preprintnumbers,amsmath,amssymb,10pt]{revtex4-1} 
\usepackage{graphicx}
\usepackage{dcolumn}
\usepackage{bm}
\usepackage{color}
\usepackage{url}
\newcommand{\kk}{\mathbf{k}}
\newcommand{\pp}{\mathbf{p}}

\newcommand{\xx}{\mbox{\boldmath $x$}}

\newcommand{\vv}{\mathbf{v}}

\begin{document}
\title{Toward an accurate Dark Matter power spectrum beyond BAO scales}
\author{Naonori S. Sugiyama}
\email{sugiyama@astr.tohoku.ac.jp}
\affiliation{Astronomical Institute, Graduate School of Science, Tohoku
University, Sendai 980-8578, Japan}  

\author{Toshifumi Futamase}
\affiliation{ Astronomical Institute, Graduate School of Science, Tohoku
University, Sendai 980-8578, Japan}

\begin{abstract}
Although there have been various theoretical studies for nonlinear evolution of Dark Matter in Newtonian gravity,
the main interest has been the nonlinear shift of Baryon Acoustic Oscillation (BAO).
In this letter, we propose a new approximated expression for nonlinear Dark Matter power spectrum applicable for much beyond BAO scales.
In particularly, the proposed expression agrees with the result of $N$-body simulation with the accuracy better than $2\%$ 
up to $k=1.0$ $[h/{\rm Mpc}]$ at $z=3.0$.
Even at $z=1.0$, the accuracy remains within $5\%$ up to $k=0.8$ $[h/{\rm Mpc}]$.
In doing so, we proved the Reg PT proposed by Bernardeau et al. (2011) 
using an approximation for the kernel functions used in the Standard Perturbation Theory, and derived a extended version of the Reg PT.
\end{abstract}


\maketitle
One of the key quantity in modern cosmology is the matter power spectrum because it contains 
many important informations on the evolution of the universe as well as of structure formation. 
Recent observation of the Baryon Acoustic Oscillation (BAO) in the power spectrum 
provides us a new method to precisely restrict cosmological parameters~\cite{Eisenstein/etal:2005}.

Recent progress of the cosmological observation greatly motivates 
various theoretical studies for accurate calculation of non-linear matter power spectrum.
At present, the most successful model is given by the Reg PT scheme~\cite{Bernardeau:2008fa,Bernardeau:2011dp}.
Especially, Taruya et al. (2012)~\cite{Taruya:2012ut} gave a theoretical prediction which agrees with $N$-body simulation very well on BAO scales
with the accuracy better than $1\%$ about up to $k=0.2$ $[h/{\rm Mpc}]$ at $z=1$.

We have also proposed a new method called the ``Wiener Hermite (WH) expansion``
which gives an accurate power spectrum over relatively wide range of wave number~\cite{Sugiyama:2012pc}.
There, we established the relation between the WH expansion and the standard perturbation theory (SPT), and showed the equivalence between 
the WH expansion and the $\Gamma$-expansion method 
\cite{Bernardeau:2008fa,Bernardeau:2011dp,Taruya:2012ut},
and derived the exponential behavior of the power spectrum in each order of the WH expansion ($\Gamma$-expansion) in terms of the SPT 
which has been known in the Renormalized Perturbation Theory (RPT) \cite{Crocce:2005xy,Crocce:2005xz,Crocce:2007dt}.

The goal of this letter is to present an accurate matter power spectrum much beyond BAO scales.
In doing so, 
we, for the first time, give a mathematical proof of the Reg PT scheme using approximated kernel functions in the SPT.
Furthermore, the approximated kernel functions allow us to extend the Reg PT 
and as a result we derive the predicted power spectrum beyond BAO scales. 

First of all, we review the standard perturbation theory (SPT).
In the situation $f=\Omega_m^{1/2}$ 
where $f \equiv d \ln D/ d \ln a $ is the growth function with $D, \ a$ 
being the growth factor and the scale factor, 
it is then known that density perturbation and the velocity divergence of Dark Matter 
may be expanded as~\cite{Bernardeau:2001qr},
\begin{equation}
		\delta(z,\kk) = \sum_{n=1}^{\infty} D^{n} \delta_{n}(\kk), \ 
		\theta(z,\kk) = -aH \sum_{n=1}^{\infty} D^n \theta_n(\kk),
		\label{}
\end{equation}
where $H$ is the Hubble parameter and $\delta_n$ and $\theta_n$ are time-independent.
The variable $\kk$ is the Fourier mode.
The divergence of Dark Matter $\theta$ is defined by $\theta \equiv \nabla \cdot \vv$ where $\vv$ is the velocity of Dark Matter.
Then, the $n$-th order solutions of the perturbation theory are given by,
\begin{eqnarray}
		\delta_n(\kk) &=&  \prod_{i=1}^{n}\int \frac{d^3p_i}{(2\pi)^3} 
		(2\pi)^3 \delta_D(\kk-\pp_{[1,n]}) F_{n}([\pp_1,\pp_n]) \delta_L(\pp_i), \nonumber \\
        \theta_n(\kk) &=&  \prod_{i=1}^{n}\int \frac{d^3p_i}{(2\pi)^3} 
		(2\pi)^3 \delta_D(\kk-\pp_{[1,n]}) G_{n}([\pp_1,\pp_n]) \delta_L(\pp_i), \nonumber \\
		\label{sol_FG}
\end{eqnarray}
where $\pp_{[1,n]}\equiv\pp_1+\cdots+\pp_n$ and $F_n(\pp_1,\dots,\pp_n) \equiv F_n([\pp_1,\pp_n])$. 
The subscript $L$ means the linearized quantity.
The kernel functions $F$ and $G$ are constructed from the mode coupling functions 
$\alpha(\kk_1,\kk_2) \equiv (\kk_1+\kk_2)\cdot\kk_2/k_1^2$ and $\beta(\kk_1,\kk_2)\equiv |\kk_1+\kk_2|^2 (\kk_1\cdot\kk_2)/2k_1^2k_2^2$
according to the recursion relation~\cite{Bernardeau:2001qr}.

Various modified perturbation theories partially sum the infinite order in the SPT, called ``resummation''.
Since the SPT has exact but formal solutions for any order of the perturbation theory using kernel functions as shown in Eq.~(\ref{sol_FG}),
any resummation theory should be rewritten in terms of the SPT.
However, there is no theory that is proved a complete relation to the SPT except for the WH expansion ($\Gamma$-expansion method).
In fact, as shown in our previous work~\cite{Sugiyama:2012pc}, 
the density perturbation $\delta$ is expanded in the WH expansion method as,
\begin{eqnarray}
		\delta(\kk) =&&  \sum_{r=1}^{\infty} \prod_{i=1}^r \int \frac{d^3k_i}{(2\pi)^3} (2\pi)^3 \delta_D(\kk-\kk_{[1,r]}) \nonumber \\
		&& \hspace{1cm} \times \delta_{\rm WH}^{(r)}([\kk_1,\kk_r]) H^{(r)}([\kk_1,\kk_r]),
\end{eqnarray}
where $H^{(i)}$ $\{i = 1,\cdots,r\}$ are stochastic variables and satisfy the conditions of Eqs.(27)-(30) in~\cite{Sugiyama:2012pc}.
The coefficients of the WH expansion ($\Gamma$-expansion) can be described using the kernel functions in the SPT as follows
($r\geq 1$):
\begin{eqnarray}
		\delta_{\rm WH}^{(r)}([\kk_1,\kk_r])
		=  \sum_{n=0}^{\infty} \delta_{r+2n}^{(r)}([\kk_1,\kk_r]),
\end{eqnarray}
where 
\begin{eqnarray}
		&& \delta_{r+2n}^{(r)}([\kk_1,\kk_r]) \nonumber \\
		&=& \frac{1}{r!}\frac{(r+2n)!}{2^n n!} \delta_1^{(1)}(k_1) \dots \delta_1^{(1)}(k_r) \nonumber \\
		&&\times  \prod_{i=1}^{n}\int \frac{d^3p_i}{(2\pi)^3} F_{r+2n}([\kk_1,\kk_r],[\pp_1,-\pp_1,\pp_n,-\pp_n])P_L(p_i). \nonumber \\
		\label{WH-r}
\end{eqnarray}
The linearized perturbation of energy density and the linear power spectrum 
are expressed by $\delta_L(\kk) = \delta_1^{(1)}(k) H^{(1)}(\kk)$ and $P_L(k) = [\delta_1^{(1)}(k)]^2$, respectively, where $|\kk| \equiv k$.
This expression in Eq.~(\ref{WH-r}) means that the density fluctuation with the $r$-the order of the WH expansion is 
the sum of all the $(r+2n)$ order of SPT. The contribution to the power spectrum from $\delta^{(r)}_{\rm WH}$ is given by,
\begin{equation}
		P_{\rm WH}^{(r)}(k) = r! \prod_{i=1}^{r} \int \frac{d^3k_i}{(2\pi)^3} 	
		(2\pi)^3 \delta_D(\kk-\kk_{[1,r]}) \left[ \delta_{\rm WH}^{(r)}([\kk_1,\kk_r]) \right]^2.
		\label{power-r}
\end{equation}
The coefficients in the $\Gamma$-expansion are simply related to ones in the WH expansion,
\begin{equation}
		\delta_{\rm WH}^{(r)} = \delta_1^{(1)}(k_1) \dots \delta_1^{(1)}(k_r)\Gamma^{(r)}(\kk_1,\dots,\kk_r).
		\label{}
\end{equation}
For the divergence of Dark Matter $\theta$, 
the same relation with $\delta$ is satisfied except for replacing of the kernel function $F$ to $G$.

Notice that once given the relation between the WH expansion and SPT,
the problem to solve the non-linear evolution of Dark Matter eventually reduces to compute the kernel functions in the SPT.
Therefore, we next consider the approximation of the kernel functions.
For $|\pp|\to0$, the kernel functions are approximated as
\begin{eqnarray}
		F_n([\pp_1,\pp_{n-1}],\pp) &\to& \frac{1}{n}\left(\frac{\pp_{[1,n-1]}\cdot \pp}{p^2} \right) F_{n-1}([\pp_1,\pp_{n-1}]), \nonumber \\
		G_n([\pp_1,\pp_{n-1}],\pp) &\to& \frac{1}{n}\left(\frac{\pp_{[1,n-1]}\cdot \pp}{p^2} \right) G_{n-1}([\pp_1,\pp_{n-1}]). \nonumber \\
		\label{ap_FG}
\end{eqnarray}
We can prove this by the induction in $n$ from the recursion relation of $F$ and $G$
using $\alpha(\pp_{[1,n]},\pp) \to 1$ and $\beta(\pp_{[1,n]},\pp) \to \pp_{[1,n]}\cdot\pp/2 p^2$.

Here, we consider the physical meaning of this approximation in Eq.~(\ref{ap_FG}).
For $|\pp_i| \to 0 \ \ \{i =2,\cdots,n\}$,
the kernel function $F$ is approximated as
\begin{equation}
		F_n([\pp_1,\pp_n]) \to \frac{1}{n!} \left( \frac{\pp_1\cdot\pp_n}{p_n^2} \right) \cdots \left( \frac{\pp_1\cdot\pp_2}{p_2^2} \right),
\end{equation}
and we derive the approximated $n$-th order solution of the perturbation theory from Eq.~(\ref{sol_FG}):
\begin{eqnarray}
		\delta_n(\kk)
		&\to& \frac{n}{n!}\int \frac{d^3p_1}{(2\pi)^3}
		\Bigg[\prod_{i=2}^{n}\int \frac{d^3p_i}{(2\pi)^3} \left( \frac{\pp_1\cdot\pp_i}{p_i^2} \right) \delta_L(\pp_i)\Bigg] \nonumber \\
		&& \times (2\pi)^3 \delta_D(\kk-\pp_{[1,n]})\delta_L(\pp_1), 
		\label{delta_n}
\end{eqnarray}
where $n$ factor is multiplied, because we choose one Fourier mode $\pp_1$ from $n$ Fourier modes $\pp_i$ $\{i = 1,\cdots,n\}$.
Since in the linearized theory the density perturbation and velocity is related as
\begin{equation}
		\frac{i \pp}{p^2} \delta_L(\pp) = \frac{\vv_L(\pp)}{aH} \equiv {\bf \Psi}_L(\pp),
		\label{}
\end{equation}
in real space Eq.~(\ref{delta_n}) becomes
\begin{equation}
		\delta_n(\xx) \to \frac{(-1)^{n-1}}{(n-1)!}({\bf \Psi}_L\cdot \nabla)^{n-1} \delta_L(\xx),
		\label{}
\end{equation}
and we finally find
\begin{equation}
		\delta(\xx) \equiv \sum_{n=1}^{\infty} \delta_n(\xx) \to \delta_{L}(\xx-{\bf \Psi}_L(\xx)).
		\label{realspace}
\end{equation}
Thus, the approximation of Eq.~(\ref{ap_FG}) extracts the effect of the coordinate transformation of the density perturbation
due to the velocity of Dark Matter.
This effect would cause the shift of the position of the BAO peak,
but hardly contribute to the non-linear evolution of Dark Matter.

Third, we give a proof of the Reg PT~\cite{Bernardeau:2008fa,Bernardeau:2011dp,Taruya:2012ut}.
Using Eq.~(\ref{ap_FG}), we can derive the approximated solution of $\delta_{r+2n}^{(r)}$ in Eq.~(\ref{WH-r}) as follows:
\begin{eqnarray}
		\delta_{r+2n}^{(r)} &\to& n \delta_{r+2n}^{(r)}\big|_{p_2,\cdots,p_n \to0}
		 - (n-1) \delta_{r+2n}^{(r)}\big|_{p_1,\cdots,p_n \to0} \nonumber \\
		 &=& \frac{n}{n!}\left( -\frac{k^2 \sigma_v^2}{2} \right)^{n-1}\delta_{r+2}^{(r)}
		- \frac{n-1}{n!}\left( -\frac{k^2 \sigma_v^2}{2} \right)^{n}\delta_{r}^{(r)}, \nonumber \\
		\label{del_r+2n}
\end{eqnarray}
where $\sigma_v^2$ is the dispersion of the velocity coming from the effect of the coordinate transformation in Eq.~(\ref{realspace})
: $\sigma_v^2\equiv \int dp P_L(p)/6\pi $.
We multiply the factor $n$ to the first term because we choose one Fourier mode from $n$ Fourier modes.
Here, we must subtract $(n-1)\delta_{r+2n}^{(r)}|_{p_1,\cdots,p_n \to0}$ because the first term in Eq.~(\ref{del_r+2n}) 
integrates the same region defined by $p_1,\cdots,p_n\to0$ $n$ times in the multiple integral in Eq.~(\ref{WH-r}).
Then, we find 
\begin{eqnarray}
		\hspace{-1cm}\delta_{\rm WH}^{(r)} &&= \sum_{n=0}^{\infty}\delta_{r+2n}^{(r)}  \nonumber \\
		&&\to \exp\left(   -\frac{k^2 \sigma_v^2}{2} \right)
		\left[ \delta_{r}^{(r)} + \left( \delta_{r+2}^{(r)} + \frac{k^2 \sigma_v^2}{2} \delta_{r}^{(r)}   \right) \right].
		\label{Reg1}
\end{eqnarray}
Similarly, we can further show the following expression:
\begin{eqnarray}
	&&	\hspace{-0.7cm}\delta_{\rm WH}^{(r)}  
		\to \exp\left(  -\frac{k^2 \sigma_v^2}{2} \right)
		\Bigg[ \delta_{r}^{(r)}+ 
		\left(  \delta_{r+2}^{(r)} + \frac{k^2 \sigma_v^2}{2}  \delta_{r}^{(r)} \right)  \nonumber \\
		&& + \left( \delta_{r+4}^{(r)} + \left( \frac{k^2 \sigma_v^2}{2}  \right)\delta_{r+2}^{(r)}
	+ \frac{1}{2} \left(\frac{k^2 \sigma_v^2}{2}  \right)^2 \delta_r^{(r)}  \right)	  \Bigg], 
	\label{Reg2}
\end{eqnarray}
These expressions of Eqs.~(\ref{Reg1}) (\ref{Reg2}) are the same with Eqs.(51) (52) in Bernardeau et al. (2011)
\cite{Bernardeau:2011dp}.
Thus, we find that the Reg PT scheme is equivalent to the approximation of the kernel functions presented in Eq.~(\ref{ap_FG}).

\begin{figure*}[t]
		\scalebox{0.46}[0.6]{\includegraphics {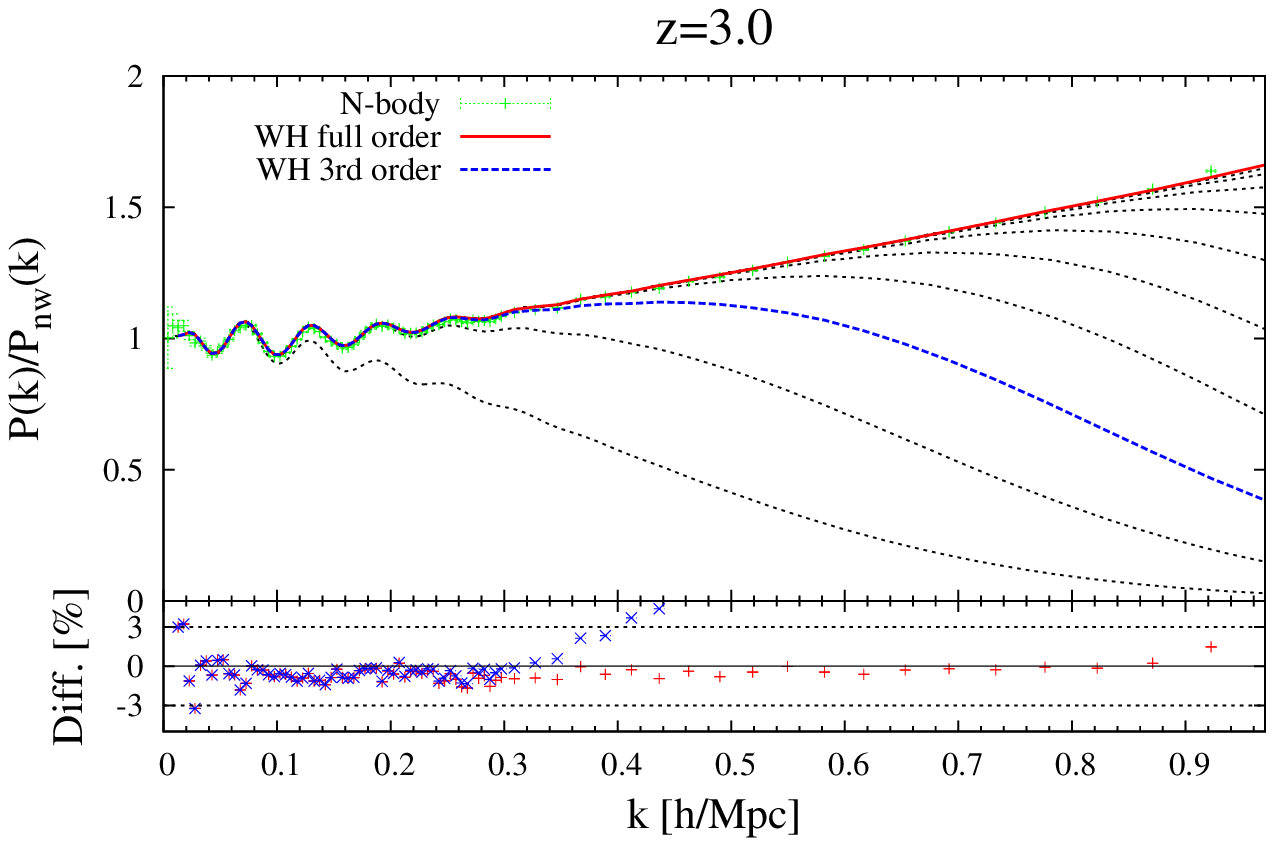}}
		\scalebox{0.46}[0.6]{\includegraphics {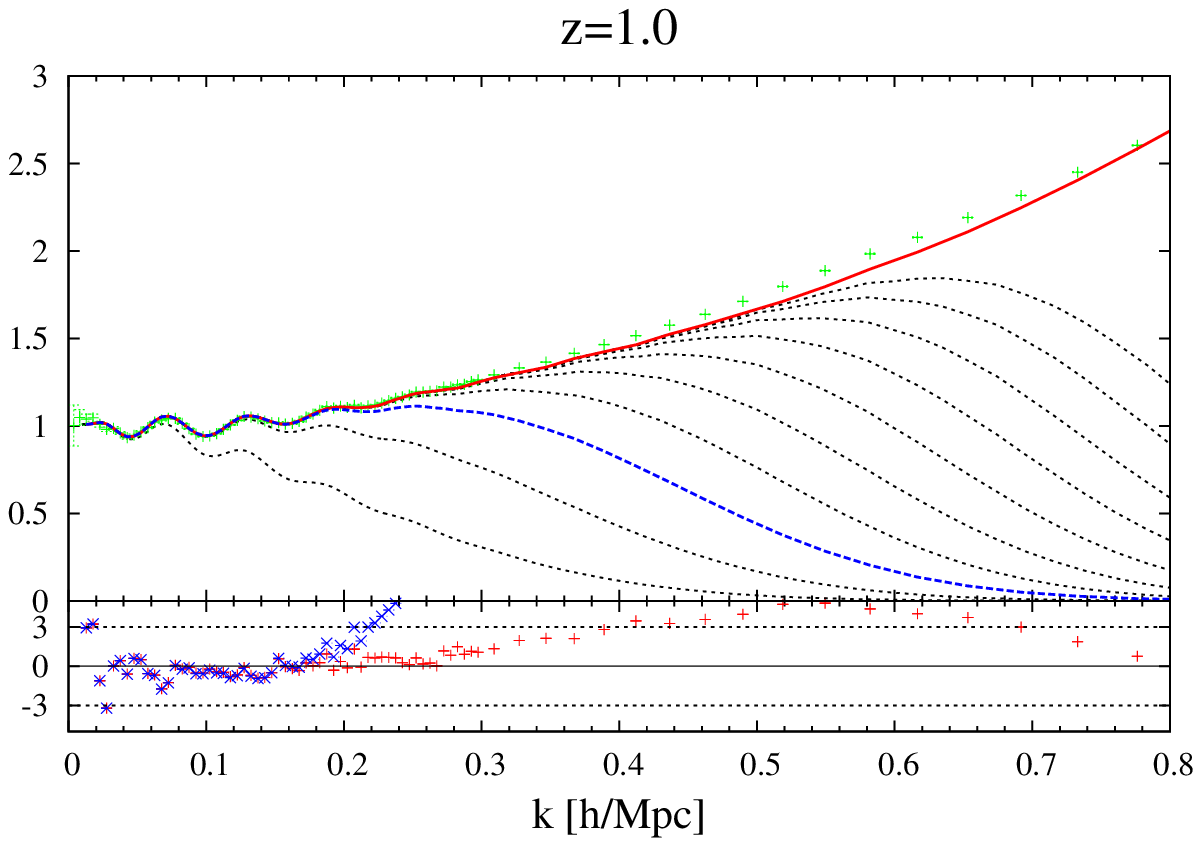}}
		\scalebox{0.46}[0.6]{\includegraphics {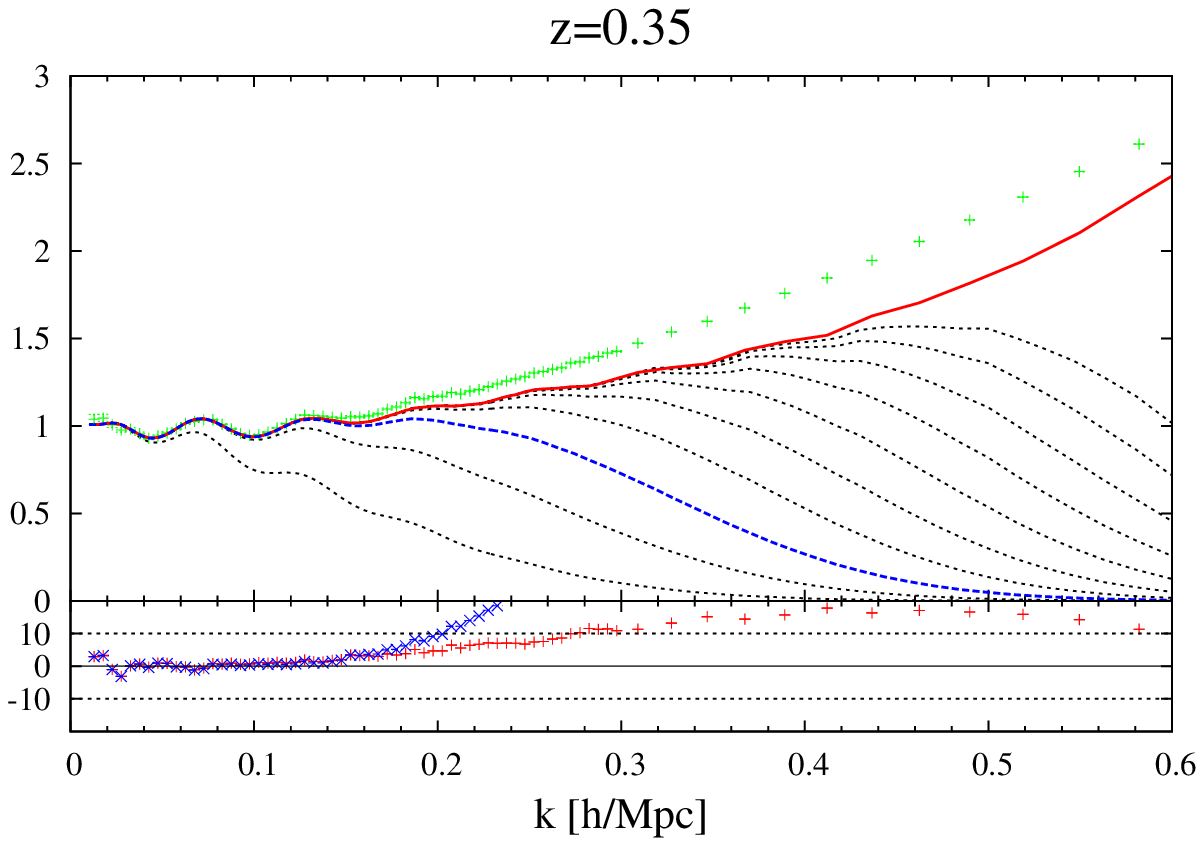}}
		\caption{The $N$-body and theoretical results are plotted for the cosmological models with the WMAP  5yr \cite{Komatsu:2008hk} 
                 ($\Omega_m = 0.279$, $\Omega_{\Lambda} = 0.721$, $\Omega_b = 0.046$, $h = 0.701$, $n_s = 0.96$ and $\sigma_8=0.817$ ),
				 at some redshifts ($z=3.0$, $1.0$ and $0.35$).
				 We show the ratio of the predicted power spectra to the smoothed reference spectra \cite{Eisenstein:1997ik},
				 $P/P_{\rm nw}$,
				 (Red solid line: the approximated full order in Eq.~(\ref{result}), blue dotted: the WH 3rd order, 
				 black dotted: each order from 1st to 10th in WH expansion, 
				 and green symbols: $N$-body results), 
				 and the fractional difference between $N$-body and analytic predicted results, 
				 $[P_{\rm Nbody}-P]/P_{\rm Nbody}$.
	             The $N$-body results used in this letter was presented by \cite{Valageas:2010yw}
\footnote{These results and their initial conditions at $z_{\rm ini} = 99$ 
were created by a public $N$-body code {\it GADGET2} and the {\it 2LPT} code, respectively \cite{Springel:2005mi,*Crocce:2006ve}.
These $N$-body simulations contains $2,048^3$ particles 
and were computed by combining the results with different box sizes $2,048 h^{-1} {\rm Mpc}$ and $4,096 h^{-1}{\rm Mpc}$, called 
$L11$-$N11$ and $L12$-$N11$.}.
We used the program presented by Taruya's homepage to compute predicted power spectra
\footnote{ \url{ http://www-utap.phys.s.u-tokyo.ac.jp/~ataruya/regpt_code.html}}.}
      \label{fig:power}
 \end{figure*}

Finally, in this letter, we consider further approximations of Eqs.~(\ref{Reg1}) (\ref{Reg2}) using Eq.~(\ref{ap_FG}).
In Taruya et al. (2012)~\cite{Taruya:2012ut}, the first and second orders of the WH expansion ($\Gamma$-expansion) 
are computed by Eq.~(\ref{Reg2}) and Eq.~(\ref{Reg1}), respectively,
and the third order of the WH expansion is evaluated just as tree level without the Reg PT scheme
(see Eq. (23)-(26) in Taruya et al. (2012)~\cite{Taruya:2012ut}).
Here, we consider any orders of the WH expansion using Eq.~(\ref{Reg2}).
When we meet incomputable terms which include the higher order of the SPT than the SPT 2-loop,
we approximate such terms using Eq.~(\ref{ap_FG}).

For the case of $r=1$ and $r=2$,
$P_{\rm WH}^{(r)}$ from Eq.~(\ref{power-r}) are given by
\begin{eqnarray}
		P_{\rm WH}^{(1)} &=&  \left[\Gamma^{(1)}_{\rm reg} \right]^2 P_L, \nonumber \\
		P_{\rm WH}^{(2)} &=&
		 2\int \frac{d^3p}{(2\pi)^3} [\Gamma_{\rm reg}^{(2)}(\pp,\kk-\pp)]^2 P_L(|\kk-\pp|) P_L(p) \nonumber \\
		 && \hspace{-1.2cm}+ k^2 \sigma_v^2 \left[ P_{\rm WH}^{(1)}-
		e^{-k^2 \sigma_v^2}
		\left( \delta_1^{(1)} + \delta_3^{(1)} + \frac{k^2 \sigma_v^2}{2} \delta_1^{(1)} \right)^2\right], \nonumber \\
		\label{eq}
\end{eqnarray}
where 
\begin{eqnarray}
		&& \Gamma^{(1)}_{\rm reg}(k) \delta_1^{(1)}(k)\equiv\delta_1^{(1)}(k)\left( 1 + \frac{k^2 \sigma_v^2}{2} +
		\frac{1}{2}\left( \frac{k^2 \sigma_v^2}{2} \right)^2 \right)  \nonumber \\
		&& \hspace{3cm} + \delta_3^{(1)}(k) \left( 1 + \frac{k^2 \sigma_v^2}{2} \right) + \delta_5^{(1)}(k), \nonumber \\
        && \Gamma^{(2)}_{\rm reg}(\kk_1,\kk_2) \delta_1^{(1)}(k_1)\delta_1^{(1)}(k_2) \equiv 
		\delta_2^{(2)}(\kk_1,\kk_2) \left(  1 + \frac{k^2 \sigma_v^2}{2} \right)  \nonumber \\
		&& \hspace{4.5cm} + \delta_4^{(2)}(\kk_1,\kk_2).
\end{eqnarray}
In Eq.~(\ref{eq}), $\Gamma_{\rm reg}^{(1)}$ and $\Gamma_{\rm reg}^{(2)}$
are the same with ones defined by Taruya et al.(2012) (see Eqs. (24) and (25) in the paper). 
We have the additional terms for $P_{\rm WH}^{(2)}$ which are approximated using Eq.~(\ref{ap_FG})
in the condition that a Fourier mode in $\delta_{\rm WH}^{(2)}(\kk_1,\kk_2)$ is close to zero: $\kk_1 \to 0$ or $\kk_2 \to 0$.
The simialr calculation for $r\geq3$ leads to the following approximation for
\begin{eqnarray}
		&&P_{\rm WH}^{(r)} \nonumber \\
		&=& \exp(-k^2 \sigma_v^2)\frac{(k^2 \sigma_v^2)^{r-3}}{(r-3)!} 
		\left[  P_{33} -k^2 \sigma_v^2 P_{22} + \frac{1}{2}(k^2 \sigma_v^2)^2P_L  \right] \nonumber \\
		&& + \exp(-k^2 \sigma_v^2)\frac{(k^2 \sigma_v^2)^{r-2}}{(r-2)!} 
		\left[  P_{22} -k^2 \sigma_v^2 P_{L}  \right] \nonumber \\
		&&  + \frac{(k^2 \sigma_v^2)^{r-1}}{(r-1)!} P_{\rm WH}^{(1)},
		\label{eq1}
\end{eqnarray}
where 
\begin{eqnarray}
		P_{22}&\equiv& 2\int \frac{d^3p}{(2\pi)^3}[F_2(\pp,\kk-\pp)]^2 P_L(p)P_L(|\kk-\pp|) \nonumber \\
		P_{33}&\equiv& 6\int \frac{d^3p_1}{(2\pi)^3}\frac{d^3p_2}{(2\pi)^3}[F_3(\kk-\pp_{[1,2]},\pp_1,\pp_2)]^2  \nonumber \\
		&& \ \ \ \ \  P_L(|\kk-\pp_{[1,2]}|)P_L(p_1)P_L(p_2).
		\label{}
\end{eqnarray}
All terms except for $\exp(-k^2 \sigma_v^2) P_{33}$ in $P_{\rm WH}^{(r)} (r\geq3)$ 
are the additional terms compared to Taruya et al. (2012).

Thus, we can compute the coefficients of the WH expansion up to any order one needs from Eq.~(\ref{eq1}).
When we summarize $P_{\rm WH}^{(r)}$ up to the infinite order,
we arrive at the following result for the approximated full power spectrum:
\begin{eqnarray}
		P(k) &=& \sum_{r=1}^{\infty} P_{\rm WH}^{(r)} \nonumber \\
		&\to& e^{k^2 \sigma_v^2} P_{\rm WH}^{(1)} + 
		(P_{22} - k^2 \sigma_v^2 P_L)\left( 1- e^{-k^2 \sigma_v^2} \right) +  \nonumber \\
		&& + \left( P_{33}- k^2 \sigma_v^2 P_{22} + \frac{(k^2 \sigma_v^2)^2}{2} P_L \right)  \nonumber \\
		&& +P_{\rm WH}^{(2)} - k^2 \sigma_v^2 P_{\rm WH}^{(1)} .
			\label{result}
\end{eqnarray}
Note that 
the behavior of our solution at low-$k$ coincide with the result in Taruya et al. (2012) and the SPT 2-loop solution.
In fact, we easily see that when we expand the exponetial factors $\exp(\pm k^2 \sigma_v^2)$,
Eq.~(\ref{result}) reduces the SPT 2-loop solution.
Furthermore, the damping factor $\exp(-k^2 \sigma_v^2/2)$ in Eqs.~(\ref{Reg1}) and (\ref{Reg2})
which comes from the random motions of Dark Matter particles in Eq.~(\ref{realspace}) 
is canceled by the correlated motion between two points of Dark Matter.
For example, it is easily visible for the first term in Eq.~(\ref{result}) $\exp(k^2 \sigma_v^2) P_{\rm WH}^{(1)}$.
Therefore, our solution in Eq.~(\ref{result}) does not have the damping behavior
unlike the result in Taruya et al. (2012).

In Fig.~\ref{fig:power}, 
we plot the predicted power spectra and $N$-body simulation results (green symbols) devided by the smoothed power spectrum without BAO
at some redshifts ($z=3.0$, 1.0, and 0.35).
The blue doted lines are the third order of the WH expansion $P_{\rm WH}^{(1)} + P_{\rm WH}^{(2)} + P_{\rm WH}^{(3)}$
corresponding to the results in Taruya et al.(2012).
Strictly speaking, we have the additional terms even for $P_{\rm WH}^{(2)}$ and $P_{\rm WH}^{(3)}$ 
but they hardly contribute to the result.
The black doted lines are $P_{\rm WH}^{(r)}$ $\{r=1,\cdots,10\}$ in the order from left to ritht,
and the red solid lines are the approximated full power spectrum in Eq.~(\ref{result}).
On BAO scales our results well agree with the $N$-body simulations as shown in Taruya et al.(2012).
Furthermore, the additional terms with the higher order of the WH expansion produce the nonlinear matter power spectrum on smaller scales.
In particular, 
at $z=3.0$ the predicted power spectrum agrees with the $N$-body result the accuracy better than $2\%$ up to $k=1.0$ $[h/{\rm Mpc}]$.
Even $z=1.0$, our solution behaves like the $N$-body result and the accuracy is kept within $5 \%$ up to $k=0.8$ $[h/{\rm Mpc}]$. 
However, at $z=0.35$ there appears the large difference between the predicted power spectrum and $N$-body result
due to the lack of taking nonlinear effects for the density perturbation.
This might be corrected if we consider the additional terms having the SPT 3-loop order.

Finally, we mention a problem of our result.
The approximation for $P_{\rm WH}^{(r)}$ $(r\geq 2)$ includes the linear power spectrum $P_L$ in Eqs.~(\ref{eq}) and (\ref{eq1}),
which leads to the BAO behavior in high-$k$ region for example slightly visible at $k=0.2-0.5$ $[h/{\rm Mpc}]$ at $z=1.0$.
However, this behavior is not real, because 
$P_{\rm WH}^{(r)}$ $(r\geq 2)$ is derived by integrating $P_L$ in Eq.~(\ref{power-r}), 
and thus the oscillatory behavior should be smoothed out.
In order to get rid of this unrealistic behavior, we would have to modify our approximate method.
We leave this as our future work.

We can also compute the correlation function from our result by truncating the WH expansion at a finite order one needs
as shown in \cite{Sugiyama:2012pc}.
We would be able to apply to the calculation of the various effects including the nonlinear evolution of the matter density perturbations
straightforwardly: redshift distortion, bispectrum, trispectrum, and bias effects etc.

We would like to thank T. Nishimichi and A. Taruya for providing us the numerical simulation results and useful comments.
This work is supported in part by the GCOE Program  ``Weaving Science Web beyond Particle-matter Hierarchy''
at Tohoku University and by a Grant-in-Aid for Scientific Research from JSPS (No. 24-3849 for NSS and No. 20540245 for TF).

\bibliographystyle{apsrev4-1}
%
\end{document}